\documentclass[12pt]{article}
\usepackage[]{epsfig}
\usepackage[centertags]{amsmath}
\usepackage{amsfonts}
\usepackage{amssymb}
\usepackage[english]{babel}
\usepackage{amsmath, amsthm, slashed,url}

\begin{document}

\title{\bf Stability analysis for soliton solutions in a gauged CP(1) theory}
\author{
Lucas Sourrouille$^a$
\\
{\normalsize \it $^a$Departamento de F\'\i sica, FCEyN, Universidad
de Buenos Aires}\\ {\normalsize\it Pab.1, Ciudad Universitaria,
1428, Ciudad de Buenos Aires, Argentina}
\\
{\footnotesize  sourrou@df.uba.ar} } \maketitle

\abstract{We analyze the stability of soliton solutions in a Chern-Simons-CP(1) model. We show a condition for which the soliton solutions are stable. Finally we verified this result numerically.}

{\bf Keywords}:CP(1) nonlinear sigma model, Gauge theory, Topological solitons

{\bf PACS numbers}: 11.10.Lm, 11.15.-q

\vspace{10mm}
\section{Introduction}
The two dimensional $CP(n)$ sigma model was introduced in the  late  seventies \cite{golo,golo1,golo2}, in the search of understanding the strong coupling effects in $QCD$. This model captures  several interesting properties, many of them present in four dimensional $QCD$\cite{witten,witten1,witten2,witten3}. Whereas in four dimensional $QCD$ is difficult to demonstrate the existence of these properties, in two dimensional $CP(n)$ sigma model it becomes comparatively simple. An important issue related to this type of models concern to the existence of soliton type solutions. For the simplest $CP(1)$ model topological solutions have been shown to exist\cite{polyakov}.  Nevertheless, the solutions are of  arbitrary size due to scale invariance. As argued originally by Dzyaloshinsky, Polyakov and Wiegmann\cite{polyakov1} a Chern-Simons term can naturally arise in this type of models and the presence of a dimensional parameter could  play some role stabilizing the soliton solutions. A first detailed consideration of this problem was done in Ref.\cite{voru} where a perturbative analysis around the scale invariant solutions (i.e no Chern-Simons coupling $\kappa=0$) showed that  the solutions were pushed to infinite size. More recently, in Ref.\cite{my3}, a nonperturbative  analysis of the solutions was done, showing that the Chern-Simons-CP(1) system, without a potential term, admits only trivial solutions in $\rm R^2$.  Nevertheless, in Ref.\cite{my5}, it was shown that the Chern-Simons-CP(1) model in absence, has a non-trivial solution if the theory is defined in ${\rm R}^2\setminus D(0,\epsilon)$, where $D(0,\epsilon)$, is a disc centered at the origin and with an arbitrary radius $\epsilon$.

This paper pretends to be a continuation of the work \cite{my3} and a generalization of the results obtained there. We will show that a  Chern-Simons-CP(1) model with a potential term, which was proposed in the reference \cite{Z}, presents a stable soliton solution if there is a critical radius $S_c$ such that the equality
\begin{eqnarray}
\int_{D_S} d^2 x \;\; B^2 = \int_{D_S} d^2 x \;\; V
\end{eqnarray}
is satisfied for all radius $S \geq S_c$, where here the subindex $D_S$ indicates that the region of integration is a disc of radius $\hat{S} $ and the letters $V$ and $B$ represent the potential term and the magnetic field.
 \vspace{5mm}
\section{The model}

We begin by considering a $(2+1)$-dimensional Chern-Simons model coupled to a complex two component field $n(x)$  described by the action

\begin{eqnarray}
S&=& S_{cs}+\int_{D} d^3 x |D_\mu n|^2 + V
\label{S1}
\end{eqnarray}
The subindex $D$ indicates that the region of integration is a disc $D$ of radius $R$\cite{my3}.
Here  $D_{\mu}= \partial_{\mu} - iA_{\mu}$ $(\mu =0,1,2)$ is the covariant derivative and $V$ is the potential term to be determined later. The term $S_{cs}$ is the Chern-Simons action given by

\begin{eqnarray}
 S_{cs}= \kappa\int_{D} d^3 x \epsilon^{\mu \nu \rho} A_\mu \partial_\nu A_\rho
\end{eqnarray}
where

\begin{eqnarray}
F_{\mu \nu}=\partial_{\mu}A_{\nu}-
\partial_{\nu}A_{\mu} \label{F}
\end{eqnarray}
The metric signature is $(1,-1,-1)$ and  the two component field $n(x)$ is subject to
the constraint $n^\dagger n = 1$.
The constraint can be introduced in the variational process with a Lagrange multiplier. Then, we extremise the following action

\begin{eqnarray}
S&=& S_{cs}+\int_{D} d^3 x \Big(|D_\mu n|^2 + V + \lambda (n^\dagger n -1)\Big)
\label{}
\end{eqnarray}
The variation of this action yields the field equations

\begin{eqnarray}
D_\mu D^\mu n -\frac{\partial V}{\partial n^\dagger} +\lambda n =0
\label{motion1}
\end{eqnarray}

\begin{equation}
\kappa\epsilon_{\mu \nu \rho} F^{ \nu \rho} = - J_\mu=i [n^\dagger D_\mu n - n(D_\mu n)^\dagger]
\label{motion}
\end{equation}
From the first of these equations we get $\lambda= -n^\dagger (D_\mu D^\mu n - \frac{\partial V}{\partial n^\dagger})$, so that

\begin{eqnarray}
D_\mu D^\mu n -\frac{\partial V}{\partial n^\dagger} = \Big(n^\dagger (D_\mu D^\mu n -\frac{\partial V}{\partial n^\dagger})\Big)n
\label{motion2}
\end{eqnarray}
The time component of Eq.(\ref{motion})
\begin{eqnarray}
2\kappa F_{12} =  -J_0 \label{gauss}
\end{eqnarray}
is  Gauss's law of Chern-Simons dynamics. Integrating it over the entire plane one obtains the important consequence that any object with charge $Q =\int_{D} d^2 x J_0$ also carries magnetic flux $\Phi = \int_{D} B d^2 x$ \cite{Echarge,E1,E2}:

\begin{eqnarray}
\Phi = -\frac{1}{2\kappa} Q,
\label{Q}
\end{eqnarray}
where in the expression of magnetic flux we renamed $F_{12}$ as $B$.

Defining the stress-tensor as\cite{Z}
\begin{eqnarray}
T_{ \mu \nu}= (D_{\mu} n)^{\dagger} D_{\nu} n +(D_{\nu} n)^{\dagger} D_{\mu} n - g_{\mu \nu} \Big( (D_\eta n)^{\dagger} D^{\eta} n - V\Big)\;\;,
\end{eqnarray}
we get

\begin{eqnarray}
E=  \int_{D} d^2 x \Big(\kappa^2 B^2 + |D_i n|^2 + V \Big) \,,
\;\;\;\;\;\
i = 1,2 \;, \label{statich}
\end{eqnarray}
which is the expression of the energy functional for the static field configuration.

Here, It is convenient to specified the potential term $V$. Following the reference \cite{Z} we define the potential as
\begin{eqnarray}
V(n)= \eta \Big(1-n^{\dagger} ( \sigma_3 n)\Big)
\end{eqnarray}
Where
\begin{eqnarray}
\sigma_3 = \ \left( \begin{array}{cc}
1 & 0  \\
0 & -1  \end{array} \right)\;\;,
\end{eqnarray}
is the third Pauli spin-matrix and $\eta$ is the coupling strength .

Let us consider, now, the following ansatz with cylindrical symmetry for the $N$ soliton solutions:

\begin{eqnarray}
n(\phi, r)=  \left( \begin{array}{c}
\cos(\frac{\theta(r)}{2})e^{i N \phi}\\
\sin(\frac{\theta(r)}{2} )\end{array} \right)
\,,
\;\;\;\;\;\
 A_\phi (r)= a(r)
\,,
\;\;\;\;\;\
A_r =0\,,
\label{ansatz}
\end{eqnarray}
In terms of this ansatz the energy (\ref{statich}) reads as
\begin{eqnarray}
E= 2\pi \int_0^R r dr \Big(\kappa^2 \left( \frac{a(r)}{r}+ \partial_r a(r) \right)^2 +\frac{1}{4}(\partial_r \theta(r))^2 \nonumber \\
+ \left(\frac{N^2}{r^2} + \frac{2Na(r)}{r}\right)\cos^2(\frac{\theta(r)}{2})+ a^2(r) + \eta(1-\cos(\theta(r))) \Big) \;,
\label{statich1}
\end{eqnarray}
whereas the field equations become

\begin{eqnarray}
\partial_r^2 a(r)+\frac{\partial_r a(r)}{r}- \frac{a(r)}{r^2} - \frac{a(r)}{\kappa^2}=\cos^2(\frac{\theta(r)}{2})\frac{N}{r\kappa^2}
\label{m1}
\end{eqnarray}
\begin{eqnarray}
r\partial_r(r\partial_r \theta(r)) + \Big(N^2 + 2Nra(r)\Big)\sin(\theta(r)) = \eta 2 r^2 \sin(\theta(r)) \;,
\label{m2}
\end{eqnarray}
In order to ensure the regularity of the field at the origin, we impose

\begin{eqnarray}
\lim_{r \to 0} \theta(r) = \pi
\,,
\;\;\;\;\;\
\lim_{r \to 0} a(r) = 0
\label{b1}
\end{eqnarray}
On the other hand, the conditions at the boundary of the disk are in principle more general. This is because the length of the radius R is arbitrary. However, if the size of the disc becomes infinite, then, we must impose boundary conditions ensuring finite energy, that is

\begin{eqnarray}
\lim_{r \to \infty} \theta(r) = 0
\,,
\;\;\;\;\;\
\lim_{r \to \infty}a(r) =-\frac{ N}{r}
\label{b2}
\end{eqnarray}
Therefore, it is convenient to use this boundary condition independently of the length of the radius. So, we fix the boundary conditions at $R$ to be

\begin{eqnarray}
\lim_{r \to R} \theta(r) = 0
\,,
\;\;\;\;\;\
\lim_{r \to R}a(r) =-\frac{ N}{r}
\label{b2}
\end{eqnarray}
this boundary conditions imply the quantization of the magnetic flux

\begin{eqnarray}
\Phi =  2\pi\int_0^R r dr\,\, \frac{\partial_r(r\,\,a(r))}{r}=-2\pi N
\label{flux}
\end{eqnarray}
If the solutions of (\ref{m1}) and (\ref{m2}) exist their scale must be set by the quantity $\kappa$. Following  Ref.\cite{mehta}, we introduce the dimensionless quantities

\begin{eqnarray}
A = \kappa a
\,,
\;\;\;\;\;\
s =\frac{r}{\kappa}
\end{eqnarray}
in terms of which (\ref{m1}) and (\ref{m2}) become

\begin{eqnarray}
\partial_s^2 A +\frac{\partial_s A}{s}- \frac{A}{s^2} - A =\cos^2(\frac{\theta}{2})\frac{N}{s}
\label{m11}
\end{eqnarray}
\begin{eqnarray}
s\partial_s(s\partial_s \theta) + \Big(N^2 + 2NsA\Big)\sin(\theta) = \eta 2 s^2 \sin(\theta)
\label{m22}
\end{eqnarray}

The energy functional (\ref{statich1})  in terms of these new variables reads as

\begin{eqnarray}
E(S)= 2\pi \int_0^S s ds \Big( \left( \frac{A}{s}+ \partial_s A \right)^2 +\frac{1}{4}(\partial_s \theta)^2 \nonumber \\
+ \left(\frac{N^2}{s^2} + \frac{2NA}{s}\right)\cos^2(\frac{\theta}{2})+ A^2  + \eta(1-\cos(\theta))\Big) \,
\;\;\;\;\;\
\label{H3}
\end{eqnarray}
For the origin we choose the following boundary conditions,
\begin{eqnarray}
\lim_{s \to 0} \theta = \pi
\,,
\;\;\;\;\;\
\lim_{s \to 0} A = 0
\label{b11}
\end{eqnarray}
while for the boundary $S=R/\kappa$ we choose,

\begin{eqnarray}
& &\lim_{s \to S}A =-\frac{ N}{S} \nonumber \\
& & \lim_{s \to S} \theta =0
\;\;\;\;\;\
\label{b22}
\end{eqnarray}

\section{Stability analysis}
In this section we analyze the stability of soliton solutions corresponding to the equations (\ref{m11}) and (\ref{m22}).

Consider the following configuration defined in the interval of length $\lambda S$

\begin{eqnarray}
\tilde{A}_{\lambda S}(s) =\frac{A_{ S}(\frac{s}{\lambda})}{\lambda}
\,,
\;\;\;\;\;\
\tilde{\theta}_{\lambda S}(s) = \theta_{S}(\frac{s}{\lambda})
\label{config}
\end{eqnarray}
Here $\lambda$ is a positive real number such that $\lambda >1$ and the configurations (\ref{config}) satisfy the boundary conditions

\begin{eqnarray}
\lim_{s \to 0} \tilde{\theta}_{\lambda S}(s) = \pi
\,,
\;\;\;\;\;\
\lim_{s \to 0} \tilde{A}_{\lambda S}(s) = 0
\label{b111}
\end{eqnarray}

\begin{eqnarray}
\lim_{s \to \lambda S}\tilde{\theta}_{\lambda S}(s)  = 0
\,,
\;\;\;\;\;\
\lim_{s \to \lambda S}\tilde{A}_{\lambda S}(s) =\frac{-N}{\lambda S}
\label{b222}
\end{eqnarray}
We can evaluate the energy functional  (\ref{H3}) for the configuration (\ref{config}) in an interval of length $\lambda S$
\begin{eqnarray}
\tilde{E}(\lambda S)= 2\pi \int_0^{\lambda S} s ds \Big( \left( \frac{\tilde{A}_{\lambda S}(s)}{s}+ \partial_s \tilde{A}_{\lambda S}(s) \right)^2 +\frac{1}{4}(\partial_s \tilde{\theta}_{\lambda S})^2 \nonumber \\
+ \left(\frac{N^2}{s^2} + \frac{2N\tilde{A}_{\lambda S}(s)}{s}\right)\cos^2(\frac{\tilde{\theta}_{\lambda S}}{2})+ \tilde{A}^2_{\lambda S}(s) + \eta(1-\cos(\tilde{\theta}_{\lambda S}))  \Big) \,
\;\;\;\;\;\
\label{H4}
\end{eqnarray}
We denote the solution corresponding to the interval $\lambda S$ as $A_{\lambda S} (s)$  and $\theta_{\lambda S} (s)$ and its energy as $E(\lambda S)$. Since the configuration (\ref{config}) satisfy the same boundary condition  as $A_{\lambda S} (s)$  and $\theta_{\lambda S} (s)$, we have that
\begin{eqnarray}
E(\lambda S) \leq \tilde{E}(\lambda S)
\label{inq11}
\end{eqnarray}
Under the transformation $s= x\lambda$ the functional (\ref{H4}) becomes
\begin{eqnarray}
\tilde{E}(\lambda S)= 2\pi \int_0^{S} x dx \Big(\frac{1}{{\lambda}^2} \left( \frac{A_{S}(x)}{x}+ \partial_x A_{S}(x) \right)^2 +\frac{1}{4}(\partial_x \theta_{S})^2 \nonumber \\
+ \left(\frac{N^2}{x^2} + \frac{2NA_{S}(x)}{x}\right)\cos^2(\frac{\theta_{S}}{2})+ A^2_{S}(x) + \eta \lambda^2(1 -\cos(\theta_{S})) \Big) \,
\;\;\;\;\;\
\label{H5}
\end{eqnarray}
Since there are evaluated in the same interval, we can compare this expression with the formula (\ref{H3}). For this purpose we can look for the values of $\lambda$ for which the equality $\tilde{E}(\lambda S)= E(S)$ is held. Using the equations (\ref{H3}) and (\ref{H5}), we obtain
\begin{eqnarray}
\int_0^{S} x dx \Big(\frac{1}{{\lambda}^2} \left( \frac{A_{S}(x)}{x}+ \partial_x A_{S}(x) \right)^2 + \eta \lambda^2(1 -\cos(\theta_{S})) \Big)
 \nonumber \\
= \int_0^{S} x dx \Big( \left( \frac{A_{S}(x)}{x}+ \partial_x A_{S}(x) \right)^2 + \eta (1 -\cos(\theta_{S})) \Big)
\,
\;\;\;\;\;\
\label{H6}
\end{eqnarray}
Renamed

\begin{eqnarray}
\lambda^2 &=& \omega
\nonumber \\[3mm]
\int_0^{S} x dx \eta (1 -\cos(\theta_{S})) &=& a
\nonumber \\[3mm]
\int_0^{S} x dx \left( \frac{A_{S}(x)}{x}+ \partial_x A_{S}(x) \right)^2 &=&b\;,
\end{eqnarray}
the expression (\ref{H6}) reduce to

\begin{eqnarray}
\omega^2 a -(a +b)\omega + b=0
\label{poly}
\end{eqnarray}
The roots of this polynomial are

\begin{eqnarray}
\omega =1
\,,
\;\;\;\;\;\
\omega = \frac{b}{a}
\end{eqnarray}
and then the possible values of $\lambda$ are

\begin{eqnarray}
\lambda =1
\,,
\;\;\;\;\;\
\lambda_{c} = \sqrt{\frac{b}{a}}
\label{val1}
\end{eqnarray}
Now, suppose that $b>a$ and choose the values of $\lambda$ satisfying $1<\lambda < \sqrt{\frac{b}{a}}$.  It is not difficult to see that for this values of $\lambda$ the following inequality is held

\begin{eqnarray}
\int_0^{S} x dx \Big(\frac{1}{{\lambda}^2} \left( \frac{A_{S}(x)}{x}+ \partial_x A_{S}(x) \right)^2 + \eta \lambda^2(1 -\cos(\theta_{S})) \Big)
 \nonumber \\
< \int_0^{S} x dx \Big( \left( \frac{A_{S}(x)}{x}+ \partial_x A_{S}(x) \right)^2 + \eta (1 -\cos(\theta_{S})) \Big)
\,
\;\;\;\;\;\
\label{}
\end{eqnarray}
and therefore

\begin{eqnarray}
\tilde{E}(\lambda S)< E(S)
\label{inq1a}
\end{eqnarray}
Comparing (\ref{inq11}) and  (\ref{inq1a}) we have that
\begin{eqnarray}
E(\lambda S)< E(S)
\end{eqnarray}
that is, the energy decreases when we enlarge the interval S. This process is valid only for $1<\lambda < \sqrt{\frac{b}{a}}$.  We can repeat the proses by considering an interval of length $\lambda_c S$ instead of the interval $S$. Now, we have

\begin{eqnarray}
\int_0^{\lambda_c S} x dx \eta (1 -\cos(\theta_{\lambda_c S})) &=& a_1
\nonumber \\[3mm]
\int_0^{\lambda_c S} x dx \left( \frac{A_{\lambda_c S}(x)}{x}+ \partial_x A_{\lambda_c S}(x) \right)^2 &=&b_1\;,
\end{eqnarray}
again, if $a_1 <b_1$, we can conclude that the energy decreases when the interval is enlarged. That is

\begin{eqnarray}
E(\lambda_{1} \lambda_c S)< E(\lambda_c S)
\end{eqnarray}
where $1<\lambda_1 < \sqrt{\frac{b_1}{a_1}}$. The process can be repeated for successive intervals provided that $b>a$ in each of this intervals. Therefore if the relation

\begin{eqnarray}
\int_0^{S} x dx \eta (1 -\cos(\theta))
<
\int_0^{S} x dx \left( \frac{A_{S}(x)}{x}+ \partial_x A_{S}(x) \right)^2
\label{inq1}
\end{eqnarray}
is held for all interval, the energy decreases indeterminably as $S\to \infty$, and thus there are no finite size soliton solution in ${\bf R}^2$.

Now, suppose
\begin{eqnarray}
\int_0^{\rho S_1} x dx \eta (1 -\cos(\theta_{\rho S_1}))
>
\int_0^{\rho S_1} x dx \left( \frac{A_{\rho S_1}(x)}{x}+ \partial_x A_{\rho S_1}(x) \right)^2
\label{inq}
\end{eqnarray}
where $S_1$ indicates the length of an arbitrary interval and $\rho$ is a real number such that $\rho >1$.
Consider the following configuration defined in the interval of length $S_1$

\begin{eqnarray}
\tilde{A}_{S_1}(s) = \rho A_{\rho S_1}(\rho s)
\,,
\;\;\;\;\;\
\tilde{\theta}_{S_1}(s) = \theta_{\rho S_1}(\rho s)
\label{config1}
\end{eqnarray}
where  $A_{\rho S_1}(s)$ and $\theta_{\rho S_1}(s)$ are the solutions of the field equations in the interval $\rho S_1$.
In virtue of equation (\ref{b11}) and (\ref{b22}), $A_{\rho S_1}(s)$ and $\theta_{\rho S_1}(s)$ must satisfied

\begin{eqnarray}
\lim_{s \to 0} \theta_{\rho S_1}(s) = \pi
\,,
\;\;\;\;\;\
\lim_{s \to 0} A_{\rho S_1}(s) = 0
\label{b1111}
\end{eqnarray}

\begin{eqnarray}
\lim_{s \to \rho S_1} \theta_{\rho S_1}(s)  = 0
\,,
\;\;\;\;\;\
\lim_{s \to \rho S_1} A_{\rho S_1}(s) =\frac{-N}{\rho S_1}
\label{b2222}
\end{eqnarray}

Then, the configuration (\ref{config1}) is subject to the following boundary conditions

\begin{eqnarray}
\lim_{s \to 0} \tilde{\theta}_{S_1}(s) = \pi
\,,
\;\;\;\;\;\
\lim_{s \to 0} \tilde{A}_{S_1}(s) = 0\;,
\label{b11111}
\end{eqnarray}

\begin{eqnarray}
\lim_{s \to S_1}\tilde{\theta}_{S_1}(s)  &=& 0
\nonumber \\
\lim_{s \to S_1}\tilde{A}_{S_1}(s)& =& -\rho \frac{N}{\rho S_1}= \frac{-N}{S_1}
\label{b22222}
\end{eqnarray}
The solutions of the field equations in the interval $S_1$, which we denote by $A_{S_1}(s)$ and $\theta_{S_1}(s)$, also satisfied the boundary conditions (\ref{b11111}) and (\ref{b22222}). Therefore, we have
\begin{eqnarray}
E(S_1) \leq \tilde{E}(S_1)
\label{inq3}
\end{eqnarray}
where $E(S_1)$ is the energy corresponding to the solution $A_{S_1}(s)$ and $\theta_{S_1}(s)$, and

\begin{eqnarray}
\tilde{E}(S_1)&=& 2\pi \int_0^{S_1} s ds \Big( \left( \frac{\tilde{A}_{S_1}(s)}{s}+ \partial_s \tilde{A}_{S_1}(s) \right)^2 +\frac{1}{4}(\partial_s \tilde{\theta}_{S_1})^2 \nonumber \\
&+& \left(\frac{N^2}{s^2} + \frac{2N\tilde{A}_{S_1}(s)}{s}\right)\cos^2(\frac{\tilde{\theta}_{S_1}}{2})+ \tilde{A}^2_{S_1}(s) + \eta(1-\cos(\tilde{\theta}_{S_1}))  \Big) \,
\;\;\;\;\;\
\label{H7}
\end{eqnarray}
Using the configuration (\ref{config1}) and under the transformation $x=\rho s$, the expression (\ref{H7}) reads as

\begin{eqnarray}
\tilde{E}(S_1)&=& 2\pi \int_0^{\rho S_1} x dx \Big( \rho^2 \left( \frac{A_{\rho S_1}(x)}{x}+ \partial_x A_{\rho S_1}(x) \right)^2 +\frac{1}{4}(\partial_x \theta_{\rho S_1})^2 \nonumber \\
&+& \left(\frac{N^2}{x^2} + \frac{2N A_{\rho S_1}(x)}{x}\right)\cos^2(\frac{\theta_{\rho S_1}}{2})+ A^2_{\rho S_1}(x) + \frac{\eta}{\rho^2}(1-\cos(\theta_{\rho S_1}))  \Big) \,
\;\;\;\;\;\
\label{H8}
\end{eqnarray}
The energy for the solutions in the interval $\rho S_1$ is

\begin{eqnarray}
E(\rho S_1)&=& 2\pi \int_0^{\rho S_1} x dx \Big( \left( \frac{A_{\rho S_1}(x)}{x}+ \partial_x A_{\rho S_1}(x) \right)^2 +\frac{1}{4}(\partial_x \theta_{\rho S_1})^2 \nonumber \\
&+& \left(\frac{N^2}{x^2} + \frac{2N A_{\rho S_1}(x)}{x}\right)\cos^2(\frac{\theta_{\rho S_1}}{2})+ A^2_{\rho S_1}(x) + \eta (1-\cos(\theta_{\rho S_1}))  \Big) \,
\;\;\;\;\;\
\label{H9}
\end{eqnarray}
Since (\ref{H8}) and (\ref{H9}) are evaluated in the same interval we can compare this formulas. In this case we look for the values of $\rho$ for which the equality $\tilde{E}(S_1)= E(\rho S_1)$ is held. Following the same steps that we do previously we arrive to the polynomial

\begin{eqnarray}
\omega^2 a -(a +b)\omega + b=0
\end{eqnarray}
However, in this case $a$ and $b$ are different from the constants present in the polynomial (\ref{poly}). In fact we have

\begin{eqnarray}
\rho^2 &=& \omega
\nonumber \\[3mm]
\int_0^{\rho S_1} x dx \left( \frac{A_{\rho S_1}(x)}{x}+ \partial_x A_{\rho S_1}(x) \right)^2&=&a
\nonumber \\[3mm]
\int_0^{\rho S_1} x dx \eta (1 -\cos(\theta_{\rho S_1})) &=& b
\end{eqnarray}

As in formula (\ref{val1}) the roots of the polynomial produce the following values of $\rho$

\begin{eqnarray}
\rho =1
\,,
\;\;\;\;\;\
\rho_{c} = \sqrt{\frac{b}{a}}
\label{val2}
\end{eqnarray}
Again, it is easy to show that choosing the values of $\rho$ satisfying the relation $1< \rho < \sqrt{\frac{b}{a}}$ we have

\begin{eqnarray}
\tilde{E}(S_1)< E(\rho S_1)
\end{eqnarray}
and therefore in virtue of (\ref{inq3})

\begin{eqnarray}
E(S_1)< E(\rho S_1)
\end{eqnarray}
We can repeat the process by choosing $\rho_c S_1$ instead of $S_1$. In that case the formulas (\ref{inq}) and (\ref{config1}) read as

\begin{eqnarray}
\int_0^{\rho_1 \rho_c S_1} x dx \eta (1 -\cos(\theta_{\rho_1 \rho_c S_1}))
>
\int_0^{\rho_1 \rho_c S_1} x dx \left( \frac{A_{\rho_1 \rho_c S_1}(x)}{x}+ \partial_x A_{\rho_1 \rho_c S_1}(x) \right)^2
\label{}
\end{eqnarray}

\begin{eqnarray}
\tilde{A}_{\rho_1 S_1}(s) = \rho_1 A_{\rho_1 \rho_c S_1}(\rho_1 s)
\,,
\;\;\;\;\;\
\tilde{\theta}_{\rho_1 S_1}(s) = \theta_{\rho_1 \rho_c S_1}(\rho_1 s)
\label{config3}
\end{eqnarray}
and finally we arrive to

\begin{eqnarray}
E(\rho_c S_1)< E(\rho_1 \rho_c S_1)
\end{eqnarray}
Of course the process can be repeated indefinitely provided that $b>a$ in all intervals. This implies that if the relation

\begin{eqnarray}
\int_0^{ S_1} x dx \eta (1 -\cos(\theta_{ S_1}))
>
\int_0^{ S_1} x dx \left( \frac{A_{ S_1}(x)}{x}+ \partial_x A_{ S_1}(x) \right)^2
\label{}
\end{eqnarray}
is valid for all interval $S_1$, the energy increases. Certainly, we have

\begin{eqnarray}
0\leq E(\rho S_1)- \tilde{E}(S_1)\leq E(\rho S_1)- E(S_1)\;,
\end{eqnarray}
where
\begin{eqnarray}
E(\rho S_1)-  \tilde{E}(S_1) = (1-\rho^2)a + (1-\frac{1}{\rho^2})b
\end{eqnarray}
The roots of this equation are
\begin{eqnarray}
\rho =1
\,,
\;\;\;\;\;\
\rho = \sqrt{\frac{b}{a}}
\label{val2}
\end{eqnarray}
Since $1<\rho<\sqrt{\frac{b}{a}}$ for all interval, the energy increases indeterminably as $S\to\infty$. The stability only can take place if $|E(\rho S_1)-  \tilde{E}(S_1)| \to 0$  and this implies that the interval $1<\rho<\sqrt{\frac{b}{a}}$ must be contracted to a point, that means $a=b$ and then $\rho=1$.

We conclude that if a soliton solution exist, then there is a critical radius $S_c$ such that the equality

\begin{eqnarray}
\int_0^{ S} x dx \eta (1 -\cos(\theta_{\hat{ S}}))
=
\int_0^{S} x dx \left( \frac{A_{\hat{ S}}(x)}{x}+ \partial_x A_{\hat{ S}}(x) \right)^2
\label{68}
\end{eqnarray}
is verified for all radius $S \geq S_c$.
\begin{figure}
\centering
\includegraphics
[height=70mm]{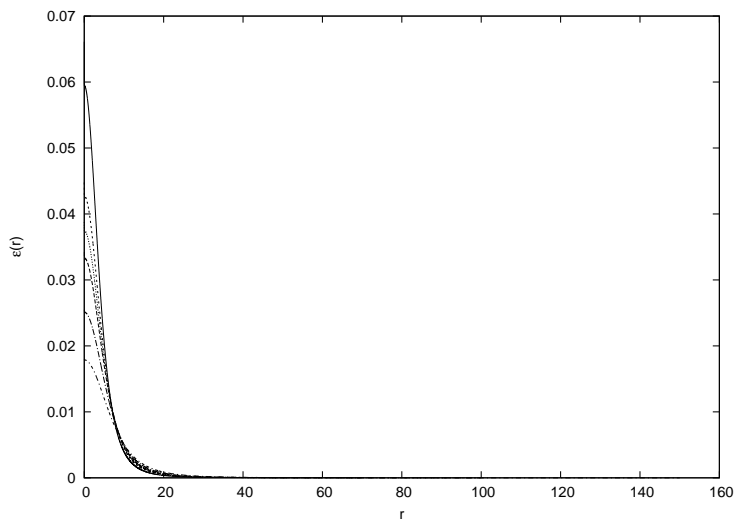}
\caption{{
The energy density, without a potential term, as a function of scaled radial coordinate $s$ for different disc sizes, from top to bottom, $S$=$30$, $50$, $60$, $70$, $100$, $150$.
}}
\label{E0}
\end{figure}
\section{Numerical Results}

\begin{figure}
\centering
\includegraphics
[height=70mm]{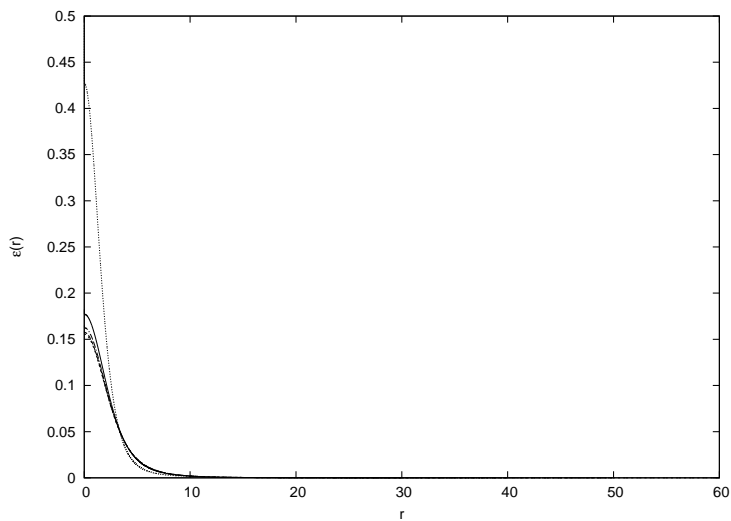}
\caption{{
The energy density, including a potential term, for different disc sizes, from top to bottom, $S$=$10$, $30$, $40$, $50$, $60$. The solution was calculated using a coupling constant $\eta =10^{-6}$.
}}
\label{E0001}
\end{figure}

The field equations (\ref{m11}) and (\ref{m22}) can be resolved numerically. From the numerical point of view one solve the field equations in a finite disc and then the analysis of the solution in an infinite plane can be carried out by increasing successively the radius of disc. In order to solve numerically the field equations for arbitrary disc sizes, the relaxation method is suitable. In Fig.\ref{E0} we show the behavior of the energy density, for $N=1$ and $S=30,50,60,70,100$, in the especial case in which there is no potential term and therefore the inequality (\ref{inq1}) is held for arbitrary disc sizes. This fact is due to the constant value of the magnetic flux, which implies $B\not=0$ and therefore $\int_{D_S} d^2 x  B^2 >0$ for all finite disc. This corresponds to problem analyzed in Ref.\cite{my3}.
The figure clearly show that as the disc size increases the energy density tends to zero and then the solution is not stable in a finite disc. These results are in concordance with our previous analysis which point out that if the inequality (\ref{inq1}) is held for arbitrary disc sizes, then the energy tends to zero.

In Fig.\ref{E0001} show the energy density, for $N=1$ and for disc sizes $S= 10,30,40,50,60$, with the inclusion of a potential term. We can see that the energy density is highly instable for small disc sizes the size (i.e. $S=10$, $20$) and becomes more stable as S increases. In fact the energy density corresponding to $S=50$ and $S=60$ are practically indistinguishable in the graph and becomes completely stabilized for $S=60$. From theoretical point of view we have that the inequality (\ref{inq1}) is held for $S<60$ and it becomes an equality for $S\geq 60$. So, $S_c=60$ is the critical radius from which the equality (\ref{68}) is verified.

\section{Conclusion}
In summary we have studied the classical solution of the Chern-Simons-CP(1) model with a potential term. Specifically we have shown that if the soliton solution exits, then there is a disc $D_{S_c}$ such that

\begin{eqnarray}
\int_{D_S} d^2 x \;\; B^2 = \int_{D_S} d^2 x\;\; V(n)
\end{eqnarray}
is satisfied for all disc $D_S \geq D_{S_c}$.
In addition we resolved numerically two situations. In the first situation we analyzed the model without a potential term, showing that the energy decreases as $S\to\infty$ , which implies the instability of the solutions. As a second case we analyzed the model with a potential term. We showed the energy decreases for $S<60$ and becomes stable for $S\geq 60$.


\begin{thebibliography}{99}
\bibitem{golo} H. Eichenherr, Nucl. Phys. B {\bf 146} (1978)
215 [Erratum-ibid. B 155 (1979) 544].
\\
\bibitem{golo1}
 V. L. Golo and A. M. Perelomov,  Lett. Math. Phys. 2, 477 (1978); Phys. Lett. B {\bf 79}, 112 (1978).
\\
\bibitem{golo2}
 E. Cremmer and J. Scherk, Phys. Lett. B {\bf 74}, 341 (1978).
\\

\bibitem{witten} A. D'Adda, M. Luscher and P. Di Vecchia, Nucl. Phys. B {\bf 146}, 63 (1978).
\\
\bibitem{witten1}
A. D'Adda, P. Di Vecchia and M. Luscher,  Nucl. Phys. B {\bf 152}, 125 (1979).
\\
\bibitem{witten2}
 E. Witten,  Nucl. Phys. B
{\bf 149}, 285 (1979).
\\
\bibitem{witten3}
R. Rajaraman, Solitons and instantons, Elsevier Science, Amsterdam, (1987). ISBN 0-444-87047-4
\\

\bibitem{polyakov}
A.A. Belavin and A.M. Polyakov, JETP Lett. {\bf 22}, 245 (1975)
\\

\bibitem{polyakov1}
I.E. Dzyaloshiskii, A.M. Polyakov and P.B. Wiegmann, Phys. Lett. A {\bf 127}, 112 (1988)
\\

\bibitem{voru} P. Voruganti, Phys. Lett. B {\bf 223} (1989) 181.
\\

\bibitem{my3} L. Sourrouille, A. Caso and G. S. Lozano, [hep-th/1002.4847], Mod. Phys. Lett. A {\bf 26}, 637 (2011) 
\\

\bibitem{my5}
Lucas Sourrouille, [arXiv:hep-th/1104.5045], Mod. Phys. Lett. A, Vol. {\bf 26}, No. 33 (2011) pp. 2523-2531.
\\

\bibitem{Z}
B.M.A.G. Piette, D.H. Tchrakian, W.J. Zakrzewski, Phys. Lett. B {\bf 339} (1994) 95.
\\

\bibitem{Echarge}
J. Schonfeld, Nucl. Phys. B {\bf 185}, 157 (1981).
\\
\bibitem{E1}
S. Deser, R. Jackiw, and S. Templeton, Phys. Rev. Lett. {\bf 48}, 975 (1982).
\\
\bibitem{E2}
S. Deser, R. Jackiw, and S. Templeton, Ann. Phys.(N.Y.) {\bf 140}, 372 (1982).
\\

\bibitem{mehta}
M.A. Mehta, J.A. Davis and I.J.R. Aitchison, Phys. Lett. B {\bf 281} (1992) 86.
\end{thebibliography}
\end{document}